# Data Visualization of Traffic Violations in Maryland, U.S.

Mumen Rababah[1], Mohammad Maydanchi[2], Shaheen Pouya[2], Mina Basiri[3], Alireza Norouzi Azad[4], Fatemeh Haji[5], Mohammad Aminjarahi[6*]

1. Department of Industrial Engineering, Faculty of Engineering the Hashemite University, P.O. Box 330127, Zarqa 13133, Jordan.  email: mumenh@hu.edu.jo

2. Department of Industrial and Systems Engineering, Auburn University, AL, United States. Emails: mzm0181@auburn.edu, shaheen.pouya@auburn.edu

3. School of Engineering, Industrial Engineering Department, Tarbiat Modares University, Tehran, Iran. Email: mina.basiri@modares.ac.ir

4. Department of Engineering Science, College of Engineering, University of Tehran, Tehran, Iran. Email: alireza.norouzi@ut.ac.ir

5. Department of Computer Engineering, Ferdowsi University of Mashhad, Mashhad, Iran. Email: fatemehaji@alumni.um.ac.ir

6. Bayes business school, City, University of London, London, United Kingdom.  Email: mohammad.aminjarahi@bayes.city.ac.uk

## Abstract

Nowadays, car use has become so common and inevitable that with a high approximation, it can be said that every family has at least one car. This has caused an increase in accidents and, subsequently, road injuries. About 1.2 million people die from road injuries yearly, and 20 to 50 million live with non-fatal injuries. Investigation of this issue is essential, considering that traffic violations have become a global concern. There, a dataset published by the Montgomery County government was analyzed using R and Python, only Maryland crimes.

The highest number of deaths is in young men, which shows an increase in traffic accident injuries in the third decade of life, and a reduction of victims of different ages was observed as a result. Factors affecting the occurrence of injuries caused by road traffic were also extracted. This can be useful in providing programs to reduce traffic violations.

**Keywords:** Traffic Violations; Data Visualization; Python; Data Analysis

# Introduction

Owning a car in the United States is popular as around 95% of American families own a car [1]. About 1.2 million people die each year from road injuries and between 20 and 50 million live with non-fatal injuries, one of the leading causes of disability for the rest of their lives. With the advancement of technology, medical care can help reduce road traffic deaths in industrialized countries [2,3].

Machines have changed the lives of individuals and communities, but their benefits have come at a price. In recent decades, the number of people killed in road accidents in high-income countries has been declining, but it cannot be ignored that the damage caused by road traffic has been increasing socially and economically for most of the world's population. In developing countries, more than 85% of all deaths and 90% of life years are associated with moderate disability due to road traffic injuries. The approach to enforcing existing laws and regulations to prevent road accidents is often inefficient and flawed [4].

# Literature Review

Road traffic accidents around the world cause severe and irreparable public health problems. Road Traffic Mortality (RTM), Road Traffic Injury (RTIs), and Disability are at the top of the list of road accident results. Traffic injuries include those injured in a car accident but have not had a disability that varies in severity. The severity of road accidents is divided into five levels according to the KABCO scale: fatal injury, debilitating injury, obvious non-debilitating injury, non-repairable injury, and financial loss. More than 1.25 million people worldwide die in road traffic accidents every year. RTIs are the leading cause of death in people aged 15 to 29. According to the World Health Organization (WHO), in 2018, road accidents will cost 3% of GDP (GPD) for most countries. If no action is taken against it, it has the potential to be included in the top 10 causes of death in the world [5,6].

The observed and predicted trend of IRTI shows that men are two to three times more affected than women because men were more active outside the home and used more means of transportation, which led to the increasing number of men in car accidents. A comparison of road traffic accident data reveals that a very small percentage of road accidents in the United States result in death, which could be due to the speed with which injuries are transported and the quality of hospital care. By examining the age range of fatalities due to road traffic accidents, the death of young people (due to injury or death in road traffic accidents) has significant social consequences [7]. Legal restrictions on using a cell phone are crucial in decreasing the number of accidents. In a study of road traffic accidents, they found that 85% of participants in the study reported using a cell phone while driving [8].

The eighth leading cause of death in the world right now is road accidents. Reducing casualties from road traffic accidents is possible by strengthening national road safety laws and implementing proven road safety interventions at the city level. Cases that lead to injuries in road traffic accidents include vigilance control, blood alcohol content, road width, lack of development of mass media campaigns for awareness, hasty driving, lack of access to victims due to non-standard roads, and the darkness of the roads. No evidence that educating teenagers reduces traffic accidents and related injuries but educating teenager drivers only leads to early licensing and may lead to an increase in the average number of teenagers involved in traffic accidents [9,10].

The drivers are the key people in road traffic accidents and driver-centered interventions [11]. Research on road structure and traffic injuries at the international level, analysis of risk factors, and interventions for road traffic and injuries worldwide is growing in recent publications. The development of motor transport leads to a parallel increase in road traffic injuries. Road injuries are a major cause of death for road users between the ages of 15 and 19. Road traffic injuries on road user groups between 10-14, 20-24, and 5-9 years old are the second and third causes of death, respectively. The most vulnerable group is young men [12].

While many drivers obey traffic signs, there is the potential for misconduct due to issues such as driver distraction and aggressive or intentional driving behaviors. Eliminating traffic violations can reduce road accidents by up to 40% [13]. The study concluded that red light cameras effectively reduced the total fatalities, but the evidence on the effect of RLCs on red light violations, total collisions, or certain types of fatalities and other violations was inconclusive. They concluded that larger studies with better control were needed, but it shows that RLCs can be reduced. Some types of traffic accidents, especially right-angle accidents, complete injured accidents, and reducing some driving violations such as speeding can be effective. Investigate the effectiveness of traffic lights and traffic regulations on traffic violations Indicates that deterrence may occur if accompanied by a violation [7,14].

There is undeniable growth in automobiles' existence, ownership, and applications these days [15]. The output of automobile ownership models has to be more detailed to address the current policy problems, including supplier selection-related problems [16]. It pertains to the segmentation of the expected automobile fleet, the segmentation of the population, as well as the need to have both short-term and long-term insight into the effect that policy initiatives would have[17,18]. Also, car ownership and vehicle type choice models are sometimes used as stand-alone models to forecast the kilometrage, fuel consumption, and emission of pollutants of the car fleet of some country or region. These models have been coupled with equations for car use (a uni-modal approach), energy use, and emissions. Existing textbook reviews of car ownership models are not very recent [19,20].

The attention to traffic concerns has been steadily improving, reflected in the rapidly expanding investment in traffic-related infrastructure. The rise in the average quality of living of the population has not only resulted in an increase in the total road distance but has also led to an increase in the number of motor vehicles. It is still an essential direction for the government and academic institutions to investigate the elements that influence the number of people who are killed or injured in road traffic accidents and to provide remedies to improve road traffic safety [12,21]. At the moment, many researchers are concentrating their efforts on investigating the elements that impact the number of individuals killed or injured in automobile accidents. These factors include people, cars, and roads. Sun et al., [22] through the regression fitting of traffic accident data, using a zero-inflated negative binomial (ZINB) regression model, found that the seniority of drivers significantly impacts the number of road traffic accidents casualties. It was discovered that gender, accident pattern, driver type, responsibility reasons, and other variables are the primary contributors to the number of people killed or injured in highway traffic accidents [23]. The amount of new road miles is shown to have a considerable influence on the number of people killed or injured in car accidents, according to the findings of several academics whose research focused on road issues. With the increase of new road mileage, the number of road traffic accident casualties will decrease significantly [10].

Researchers have examined various aspects of road traffic accident casualties and road traffic safety. On the other hand, from the point of view of the effect of traffic accident casualties, it is primarily important to investigate the influence of people, cars, and roads on traffic accident casualties in an isolated fashion. At the same time, there is a paucity of research on the incremental influence of many variables. There are few methods to think about the economic-road-population aspects [24]. When it comes to the topic of road traffic safety, the focus is primarily on the macro level; policy recommendations are only based on theoretical analysis; there is a lack of effective data support, and there is very little research on road traffic safety that is conducted through data analysis related to the number of people killed or injured in traffic accidents [9,25].

The actions of drivers have a significant impact on the safety of the traffic. In recent years, the percentage of road traffic accidents that were the consequence of human causes has been as high as 80 percent to 90 percent, as shown by the data about traffic accidents that have been collected from different nations [26,27]. Gilandeh et al. recruited forty drivers to utilize a driving simulator and drive in a variety of settings so that they could determine the human elements that contribute to unsafe driving behaviors [28]. According to different findings, males are sometimes blamed to be responsible for a greater share of the road accidents caused by speeding than women, although in many researches there is no significant difference in the number of accidents caused by male or female drivers [29,30]. There was a link between hazardous behaviors and accidents, which suggests that initiatives to promote corporate safety culture may decrease risky driver behaviors. There was also an indication of a positive association between unsafe behaviors and accidents [31]. When Chuang and Wu tested the stressors of Taiwanese bus drivers using the effort-reward imbalance model (E.R.I.) and the universal ERI scale, they discovered that the primary stressors for professional drivers were physical demands, overtime, and stress-induced sleep problems [32].

It has long been established that drinking alcohol reduces one's ability to drive safely and raises the likelihood of being involved in an accident. Research has shown that when a person is operating a vehicle under the influence of alcohol, the probability of being involved in an accident that results in serious injury or death is high [33]. It is estimated that drunk driving claims the lives of 10,000 people annually in Europe [34]. Accidents caused by drivers under the influence of alcohol account for around 31 percent of all road deaths in the United States [35]. Driving under the influence of alcohol almost always results in a devastating accident. Even if just a trace quantity of alcohol is assumed, drivers have a more significant than twofold increased risk of being involved in a traffic collision compared to sober drivers [36]. Because of this, several nations have spent a significant amount of time and effort working on solutions to the problem of drunk driving over a significant length of time. These solutions include advertising, education, and strict legislation against drunk driving. Laws have been passed that make it illegal to get behind the wheel after consuming alcoholic beverages, and those who do so face harsh penalties [6]. Between 0.01 percent and 0.08 percent is the legal limit for blood alcohol content (BAC). For instance, the limit in Sweden is 0.02 percent, whereas the restriction in Israel, Korea, and Australia is 0.05 percent, and the limit in Canada, England, Mexico, and the United States is 0.08 percent. Drinking and driving are illegal in China if the driver has a blood alcohol concentration (BAC) of 0.02 percent or more, and the motorist will be punished for this offense. In addition, driving with a blood alcohol concentration (BAC) of 0.08 percent or greater is regarded as driving under alcohol, which is a criminal offense [37].

## Methodology

The dataset that has been selected for this project was published by Montgomery County Government [38], which includes more than a million records with 35 attributes. These attributes vary between stop description date and time, race, gender, car model, locations, etc.

The analysis was done using R and python by asking several research questions to understand the relationship between different fields. Before dealing with the above step, a checking process took place on the data to determine if it needed to be cleaned first. Based on the majority of data, only the violations in Maryland (87.3% of total data) were analyzed. The following graphs and figures were generated from in-depth data analysis.

# Results

As could be seen in Figure 1 that when a driver failed to obey properly to instructions that are on a placed traffic control device was ranked as the most frequent cause for being pulled over among more than 9000 different stops causes, and a simple example for this one is that when a driver ignores no turn left sign.

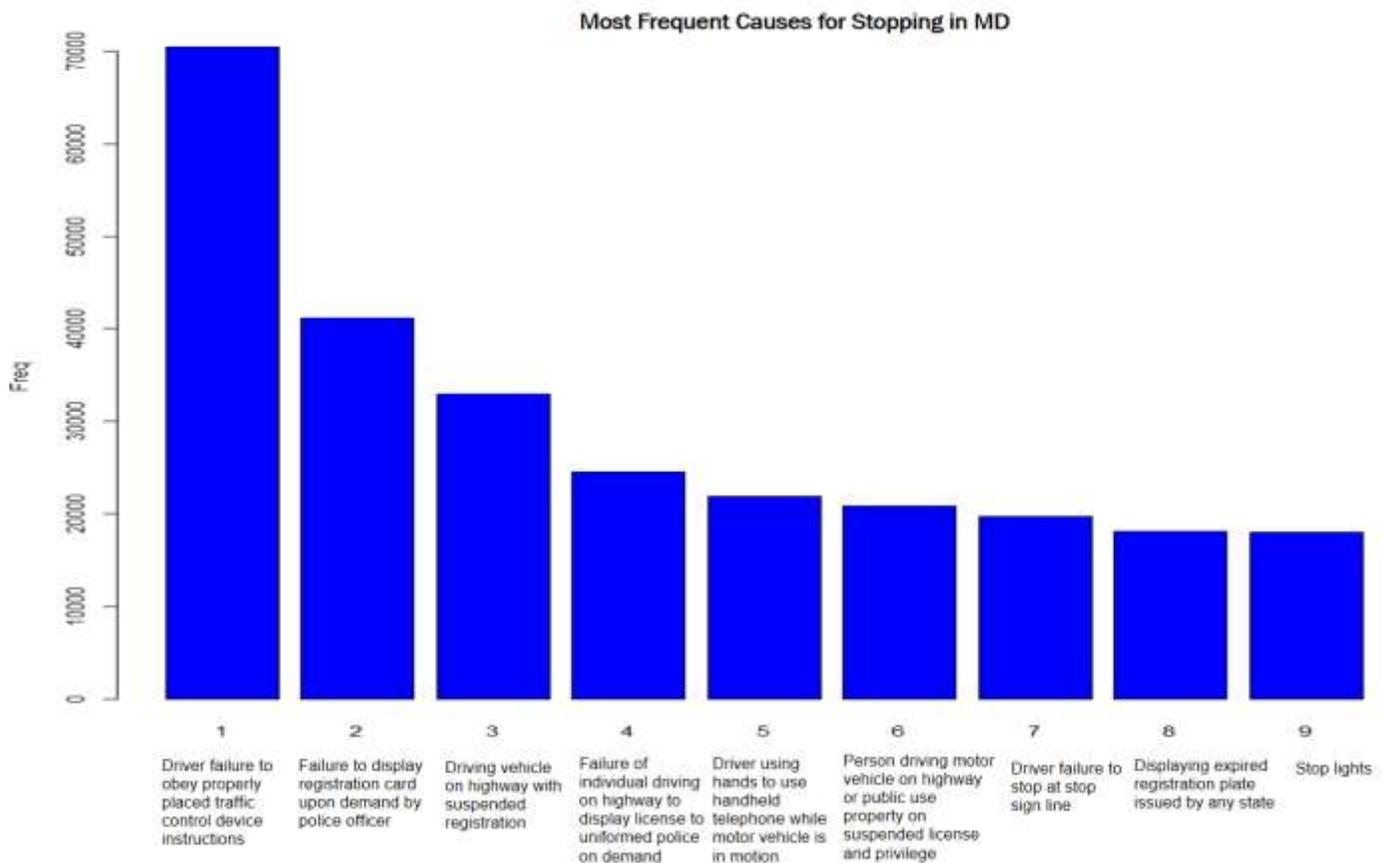

*Figure 1: Most frequent causes for stopping in MD.*

More deep analysis regarding time and date was done by analyzing the frequency of stops during each season, as shown in Figure 2. As can be seen in the graph, that winter has 229,998 stops, spring has 215,492 stops, summer has 240,674 stops, and fall has 228,868 stops. So, based on this, a conclusion could be built that getting stops during summer is more probable. Also, the frequency of stops for each month was calculated, and Figure 3 describes the distribution of stops based on months. The significant finding was that the number of stops was the minimum in June.

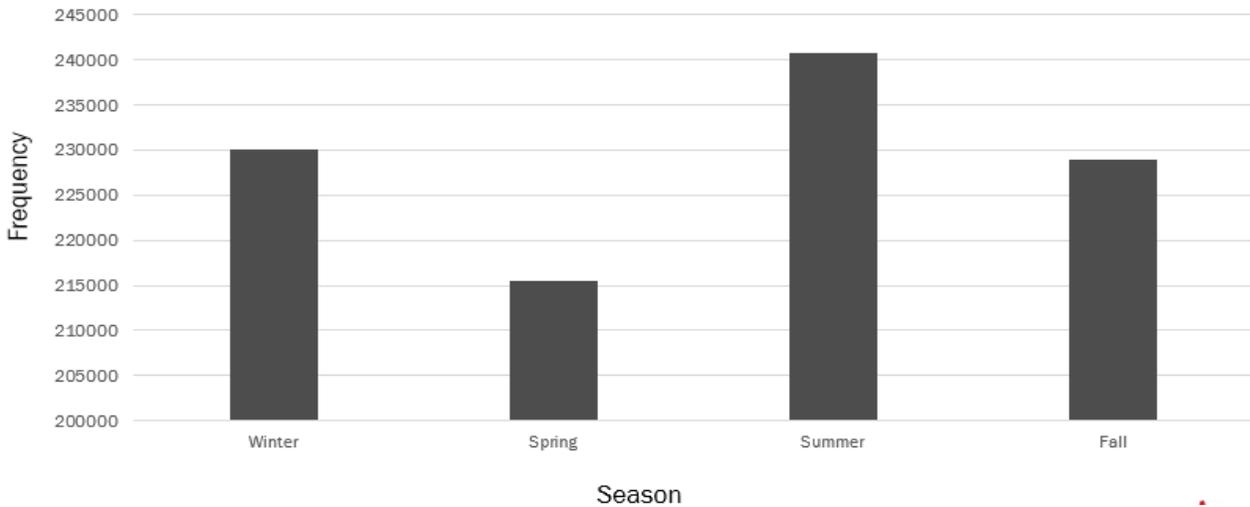

*Figure 2: Distribution of stops based on season in MD.*

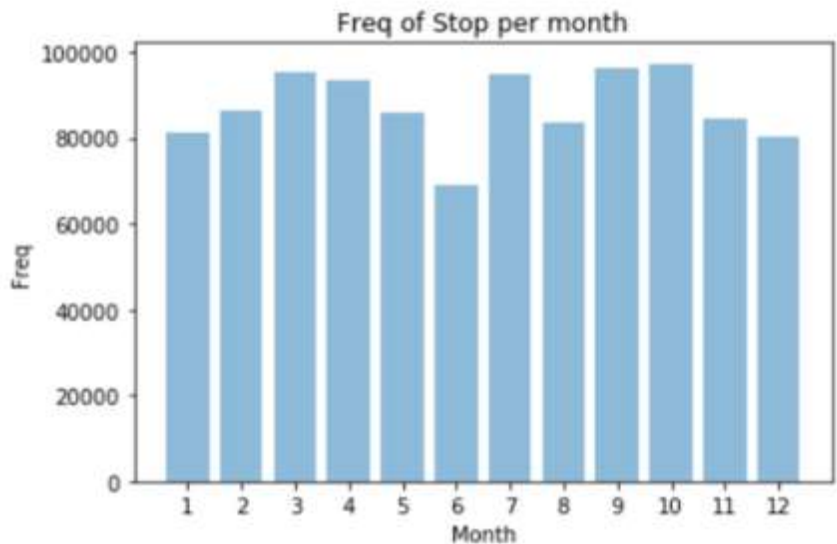

*Figure 3: Frequency of stops per month in MD.*

Moreover, knowing which hours of the day have gotten the most stops were questionable. Figure 5 shows the traffic load per hour [39]. These percentages are the rate of stops (relatively to the cars on the streets) per time of day. Notice that we assumed 6 am -12 pm as the morning, 12-6 pm as afternoon, 6 pm -12 am as night, and 12-6 am as midnight. Based on the pie chart, the lowest number of stops was in the mornings with 11%. Also, 17% of stops occurred during afternoons. Nevertheless, most of the stops in Maryland occurred during the midnights. It is reasonable as, during the night, it is possible for drivers not to see signs or be more tired.

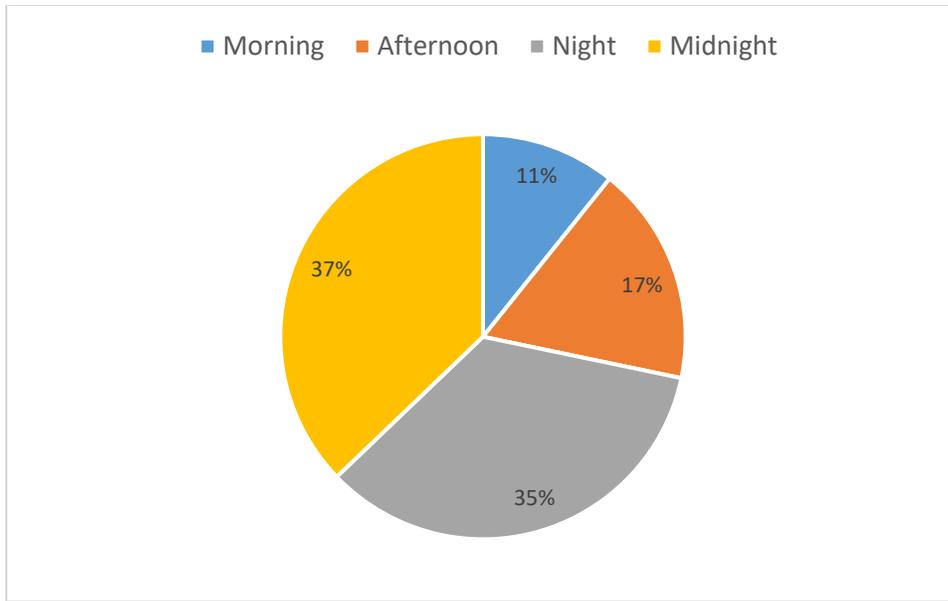

*Figure 4: Percentage of traffic violations based on the hours of the day*

One of the essential questions that a lot of people is asking is if there is a relationship between alcohol and different attributes such as fatal, personal injuries, and properties damaged, and the results were based on Table 1, 2, and 3 that alcohol does not have a significant effect on these attributes.

Table 1: Effect of alcohol on fatal.

| Alcohol vs fatal | No | Yes |
|---|---|---|
| No | 913271 | 212 |
| Yes | 1549 | 0 |

Table 2: Effect of alcohol on personal injuries.

| Alcohol vs personal injury | No | Yes |
|---|---|---|
| No | 902246 | 11237 |
| Yes | 1508 | 41 |

Table 3: Effect of alcohol on properties damage.

| Alcohol vs property damage | No | Yes |
|---|---|---|
| No | 894856 | 18627 |
| Yes | 1391 | 158 |

# Conclusion

According to this survey, the age group of young males had the most prominent casualties. The third decade of life is associated with an increased risk of being injured in a car accident. It demonstrates that persons of age groups known to be the most active and productive are nonetheless engaged in road traffic injuries, which add severe economic losses to society. According to the findings of this research, the percentage of people who were victims dropped significantly after the age of 60 and before the age of 14. It's possible that the cause is convenient access to automobiles. Living in a remote area, lacking information about existing regulations, or inadequate enforcement of those laws are all possible reasons.

In the current investigation, almost one-third of the drivers had ingested alcohol at some point. The incidence of road traffic injuries is influenced by a wide variety of human and environmental risk factors, including but not limited to age, addiction to alcohol, not having a driver's license, the kind of vehicle, etc. It is possible to avoid fatalities and problems if we exercise proper control over these parameters.

In order to expand the traffic study scenario, future research can focus on the following topics. First, analyze this traffic data in such a way as to find an optimal solution for the problem. Second, analyze the influence of increasing the number of road lanes on the actions of car drivers and consequently on traffic violations.